\documentclass[twocolumn,floatfix,prb,aps,showpacs]{revtex4}
\usepackage{graphicx,amsmath,amssymb}

\begin{document}

\newcommand{\CORR}[1]{\textbf{#1}}

\title{Theoretical study of even denominator fractions in graphene: Fermi sea versus paired states of composite fermions}
\author{Csaba T\H oke$^{1,2}$ and J.~K.~Jain$^{1}$}
\affiliation{$^{1}$Department of Physics, 104 Davey Lab, Pennsylvania State University, University Park PA, 16802}
\affiliation{$^{2}$Institut f\"ur Theoretische Physik, Johann Wolfgang Goethe-Universit\"at, 60438 Frankfurt/Main, Germany}
\date{\today}

\begin{abstract}
The physics of the state at even denominator fractional fillings of Landau levels depends on the Coulomb pseudopotentials,
and produces, in different GaAs Landau levels, a composite fermion Fermi sea, a stripe phase, or, possibly, a paired composite fermion state.  We consider here even denominator fractions in graphene, which has different pseudopotentials as well as a possible four fold degeneracy of each Landau level.  We test various composite fermion Fermi sea wave functions (fully polarized, SU(2) singlet, SU(4) singlet) as well as the paired composite fermion states in the $n=0$ and $n=1$ Landau levels and predict that
(i) the paired states are not favorable, (ii) CF Fermi seas occur in both Landau levels, and (iii) an SU(4) singlet composite fermion Fermi sea is stabilized in the appropriate limit. The results from detailed microscopic calculations are generally consistent with the predictions of the mean field model of composite fermions.

\end{abstract}

\maketitle

\section{Introduction}

Although the fractional quantum Hall effect\cite{Tsui82} (FQHE) has not yet been observed in graphene, several papers have 
studied this possibility theoretically both in the SU(2) limit\cite{FQHEgraphene} (when the Zeeman splitting is not small but the
valleys are degenerate) and in the SU(4) limit\cite{graphenesu4} (when both spins and valleys are degenerate).
These studies show that, as for GaAs, the FQHE states are well described by the composite fermion (CF) theory\cite{CF} and occur at the filling factors
\begin{equation}
\nu^{(n)}=\frac{m}{2pm\pm1}
\end{equation}
where $\nu^{(n)}$ is the partial filling of electrons (corresponding to total filling of $\nu=4n-2+\nu^{(n)}$) or holes\cite{holes} (at $\nu=4n+2-\nu^{(n)}$) in the graphene Landau level with index $n$, $2p$ is the CF vorticity, and $m$ is the number of filled $\Lambda$ levels (also known as CF Landau levels). There are differences, however.  While FQHE in GaAs is much stronger in the lowest Landau level (LL), the FQHE in graphene is expected to be as strong in the $|n|=1$ Landau level as in the $n=0$ LL\cite{FQHEgraphene}.  More interestingly, many new incompressible CF states become possible because of the SU(4) symmetry\cite{graphenesu4}.

This work addresses the nature of the state at $\nu^{(n)}=\frac{1}{2p}$. If the model of weakly interacting composite fermions remains valid in the limit of $m\rightarrow \infty$, then we expect a Fermi sea of composite fermions. In GaAs, the fully spin polarized Fermi sea of composite fermions has been extensively studied\cite{FStheory} and confirmed\cite{FSexp} at $\nu=1/2$,
and good evidence exists for a spin-singlet CF Fermi sea (CFFS) in the limit of vanishing Zeeman energy\cite{Park,Park2}.  At $\nu=5/2$ in the second ($n=1$) Landau level, it is currently believed, although not confirmed, that the residual interactions between composite fermions produce a p-wave paired state of composite fermions, described by a Pfaffian wave function\cite{MRGWW}.  In still higher Landau levels an anisotropic stripe phase is believed to occur.

CF Fermi sea is an obvious candidate at half fillings in graphene, although it will have a richer structure associated with it.  In the SU(4) symmetric limit, the mean field model of composite fermions predicts an SU(4) singlet CF Fermi sea, which has no analog in GaAs. The p-wave paired state of composite fermions is also a promising candidate, especially at $\nu^{(1)}=1/2$ in the $n=1$ LL, and it is interesting to ask if the graphene Coulomb matrix elements can make it more stable than the standard GaAs Coulomb matrix elements. For completeness, we also consider a so-called hollow-core state\cite{HR} describing the spin-singlet pairing of composite fermions, and, as in GaAs\cite{Park}, find it not to be relevant.  We note that our $n=0$ Landau level results below, as well as in Ref. \onlinecite{graphenesu4}, also apply to the CF physics in valley degenerate semiconductor systems \cite{Shayegan}.

\section{Model}

The low-energy states of graphene are described in the continuum approximation by a massless Dirac Hamiltonian\cite{Semenoff}
\begin{equation} 
H^{\text{gr.}}=v_F
\begin{pmatrix}
\vec\sigma\cdot\vec\Pi & 0\\
0 & (\vec\sigma\cdot\vec\Pi)^{\textrm{T}}\\
\end{pmatrix} + \Delta P_z + g\mu_B \vec B\cdot\vec S,
\end{equation} 
that acts on a 4-spinor Hilbert space.
Here $\vec S$ denotes the spin and $\vec P$ the pseudospin associated with the valley degree of freedom,
$v_F\approx 10^6$ m/s is the Fermi velocity, $\vec\Pi=\vec p+\frac{e}{c}\vec A$,
and $\Delta$ is the on-site energy difference between the two sublattices.
The single particle spectrum of $H^{\text{gr.}}$ is
\begin{equation} 
E_{nps}={\rm sgn}(n)\sqrt{\frac{2\hbar v_F^2 eB|n|}{c}} + \Delta p +  g\mu_B Bs,
\end{equation} 
where $s,p\in\left\{\frac{1}{2},-\frac{1}{2}\right\}$ are the eigenvalues of $S_z$ and $P_z$, respectively,
and $n$ is the Landau level index.
In the limit $g\to0,\Delta\to0$ each Landau level is 4-fold degenerate, giving rise to an SU(4) internal symmetry.
We consider below only the SU(4) symmetric part of the Hamiltonian explicitly; from these results, 
the energy of any given wave function in the presence of certain kinds of symmetry 
breaking terms (for example, the Zeeman coupling) can be obtained straightforwardly, and level crossing transitions 
as a function of $g$ and $\Delta$ can be obtained.
The conditions for SU(4) symmetry have been discussed in Refs.\ \onlinecite{graphenesu4}
and \onlinecite{Goerbig}.

Because we are interested in bulk properties, we will use the spherical geometry, in which electrons move on the
surface of a sphere and a radial magnetic field is produced by
a magnetic monopole of strength $Q$ at the center.\cite{Haldane,Fano}
Here $2Q\phi_0$ is the magnetic flux through the surface of the sphere; $\phi_0=hc/e$, and
$2Q$ is an integer according to Dirac's quantization condition.

The interelectron interaction is conveniently parametrized in terms of pseudopotentials\cite{Haldane} $V_m$, where $V_m$ is the energy of two electrons in relative angular momentum state $m$.
The problem of interacting electrons in the $n$-th LL of graphene can be mapped
into a problem of electrons in the $n=0$ LL with an effective interaction that has pseudopotentials\cite{FQHEgraphene,Nomura} 
\begin{equation} 
\label{effective}
V_m^{(n)\textrm{gr.}}=\int\frac{d^2k}{(2\pi)^2}\frac{2\pi}{k}F_n(k)e^{-k^2}L_m(k^2),
\end{equation} 
where the form factor $F_n$ is
\begin{equation}
F_0(k)=1,\quad F_n(k)=\frac{1}{4}\left(L_n\left(\frac{k^2}{2}\right) + L_{n-1}\left(\frac{k^2}{2}\right)\right)^2.
\end{equation}
For an evaluation of the energies of various variational wave functions by the Monte Carlo method, we need the real-space interaction.  In the $n=0$ LL this interaction is simply $V(r)=1/ r$, where  $r$ is taken as the chord distance in the spherical geometry.
In other Landau levels we use an effective real-space interaction in the lowest Landau level that reproduces
the higher Landau level pseudopotentials in Eq.~(\ref{effective}).
We determine such an effective real space interaction in the \emph{planar} geometry, and use it on the sphere. This procedure is exact in the thermodynamic limit, and it is usually reasonable also for finite systems. Following Ref.\ \onlinecite{graphenesu4}, in the $|n|=1$ LL we use the form
\begin{equation} 
V^{\text{eff}}(r)=\frac{1}{r}+\sum_{i=0}^M c_i r^i e^{-r}.
\end{equation} 
The coefficients $\{c_i\}$ are given in Ref.\ \onlinecite{graphenesu4}.
We will assume parameters such that the finite thickness of the 2DEG and Landau level mixing have negligible effect.

To build composite fermion trial wave functions, we will use the following consequence of Fock's cyclic condition\cite{graphenesu4}. The orbital part of one member of the SU($n$), namely the highest weight state, can be constructed as
\begin{equation}
\label{orbital}
\Phi=\mathcal{P}_{\text{LLL}}\Phi_1\Phi_2\cdots\Phi_{n} \prod_{i<j}(u_iv_j-u_jv_i)^{2p},
\end{equation}
where $\Phi_s$'s are Slater determinants such that any state $(n,m)$ in $\Phi_s$ is also filled in $\Phi_{s-1}$
(conversely, if $(n,m)$ is empty in $\Phi_s$, then it is also empty in $\Phi_{s+1}$);
$\mathcal{P}_{\text{LLL}}$ is the projection into the lowest ($n=0$) Landau level\cite{projection};
and the last factor, the Jastrow factor, attaches $2p$ vortices to each fermion.
Here $u_i=\cos\left(\theta_i/2\right)e^{-i\phi_i/2}$, and $v_i=\sin\left(\theta_i/2\right)e^{i\phi_i/2}$.
The complete wave function is
\begin{equation} 
\Phi'(\{\vec r_j\})={\cal A}\left(
\Phi(\{\vec r_j\})\prod_{t=1}^n\prod_{i=\min_t}^{\max_t}\alpha^t_i
\right),
\end{equation} 
where $\{\alpha^t\}$ is a basis of the ($n$-dimensional) fundamental representation of SU($n$),
$M_t$ is the number of particles in the $\alpha^t$ state,
$\min_1=1,\max_1=M_1,\min_2=M_1+1,\max_2=M_1+M_2,\dots$, and $\cal A$ is the antisymmetrizer.

We define the CF Fermi sea as the thermodynamic limit of an integral number of filled Landau
levels at an effective monopole strength $q=0$ for composite fermions.  Clearly, if $\Phi_1,\dots,\Phi_n$ are identical, then Eq.~(\ref{orbital}) yields a legitimate trial wave function. We will label this state ``CFFS $[\frac{N}{n},\dots,\frac{N}{n}]$."  As the effective monopole strength of composite fermions $q$ is related to the real monopole strength $Q$ as
\begin{equation} 
Q=q+p(N-1),
\end{equation} 
the filling factor is, assuming $q=\mathcal O(1)$,
\begin{equation} 
\nu^{(n)}=\lim_{N\to\infty}\frac{N}{2Q+1}=\lim_{N\to\infty}\frac{N}{2p(N-1)+1}=\frac{1}{2p}.
\end{equation} 

The Pfaffian wave function \cite{MRGWW}, which is 
one of the candidates for the FQHE state at $\nu=\frac{5}{2}$ in GaAs samples\cite{Willett1}, has the form
\begin{equation} 
\Psi^{\text{Pfaff}}_{1/2p}=\text{Pf}\left(\frac{1}{u_iv_j-v_iu_j}\right)\prod_{i<j}(u_iv_j-u_jv_i)^{2p}.
\label{Pfaff1}
\end{equation} 
on the sphere. By assumption, the Pfaffian wave function uses one spin band only.  
We also consider the hollow-core state \cite{HR} 
\begin{equation} 
\Psi^{\text{hollow-core}}_{1/2p}=\textrm{det}\left(M_{ij}\right)\prod_{i<j}(u_iv_j-u_jv_i)^{2p},
\label{Hollow1}
\end{equation} 
where $M_{ij}=(u_iv_{i+N/2}-u_{i+N/2}v_i)^{-2}$ is an $\frac{N}{2}\times\frac{N}{2}$ matrix.
This state is a spin singlet in the system with SU(2) symmetry; its symmetry becomes SU(2)$\times$SU(2) in the SU(4) symmetric limit.
Because of the last factor in Eqs. (\ref{Pfaff1}) and (\ref{Hollow1}), which converts electrons into composite fermions, these wave functions describe paired states of composite fermions.

\section{Results and conclusions}

\begin{figure*}[!htbp]
\begin{center}
\includegraphics[width=0.9\textwidth, keepaspectratio]{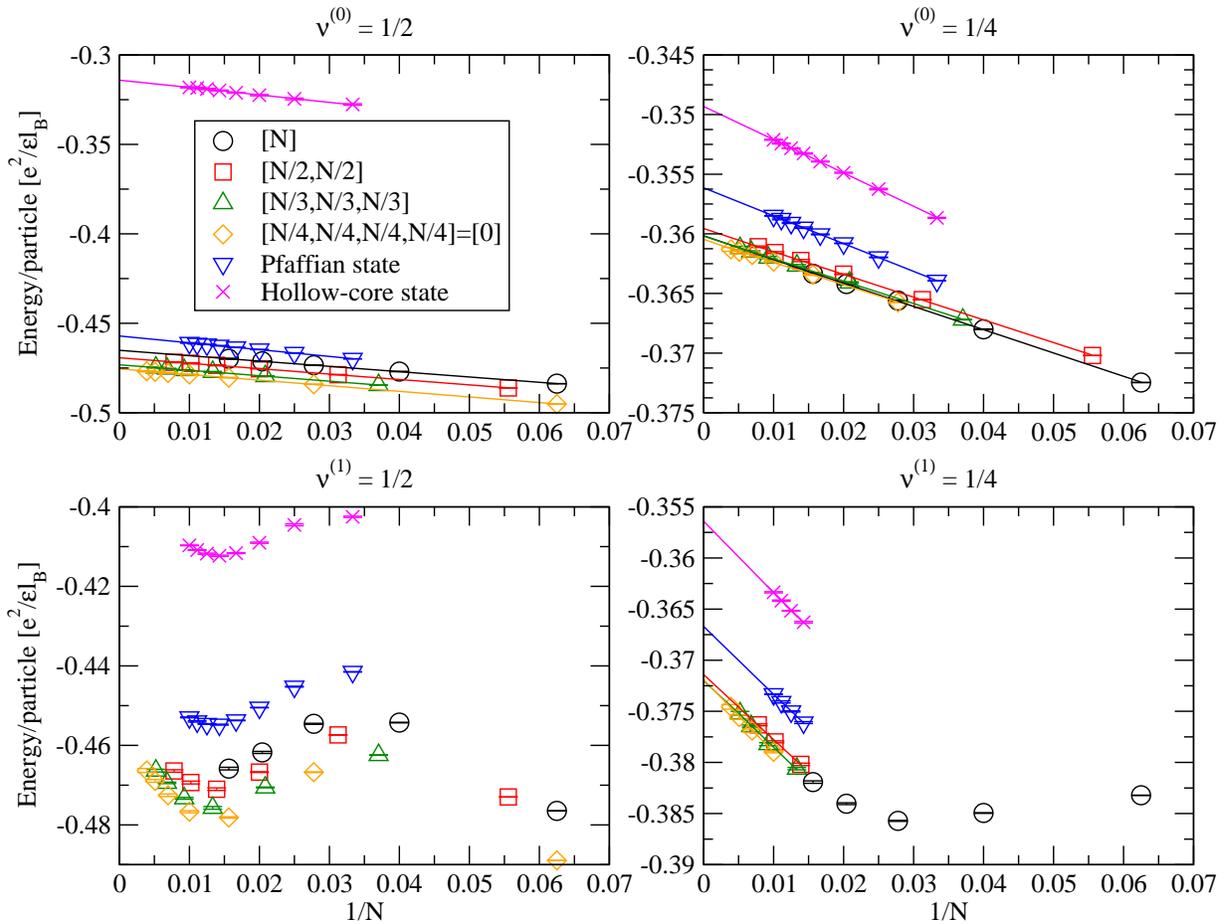}
\end{center}
\caption{\label{fs} (Color online)
Energy per particle, in units of $e^2/\epsilon l$,  for several wave functions (four CF-Fermi-sea states, the Pfaffian wave function, and the hollow core wave functions) at $\nu^{(n)}=\frac{1}{2}$ and $\frac{1}{4}$ in the $n=0$ Landau level (top) and in the $|n|=1$ Landau level (bottom).  Statistical error from Monte Carlo sampling is also shown.  Here $\epsilon$ is the background dielectric constant and $l$ is the magnetic length.  Extrapolation to the thermodynamic limit is given, wherever possible; the thermodynamic energies are quoted in Table \ref{energies}.
}
\end{figure*}

We have studied CF Fermi sea states containing as many as 256 composite fermions (64 particles per Landau band),
and our principal  results\cite{back} are given in Fig.~\ref{fs} and Table \ref{energies}.
These pertain to fillings $\nu^{(0)}=1/2$ ($\nu=\pm 3/2$);  $\nu^{(0)}=1/4$ ($\nu=\pm 7/4$);  $\nu^{(1)}=1/2$ ($\nu=\pm 5/2,\; \pm 11/2$);
$\nu^{(1)}=1/4$ ($\nu=\pm 9/4,\; \pm 23/4$).
(In relating $\nu^{(n)}$ to $\nu$, we have included the possibility of forming the state from either electrons or holes in a given Landau level.)
When the spin or valley degeneracy is broken, the above study applies to many other half integral states also.
To obtain the energy of the CF Fermi sea, we consider finite systems at $B^*=0$ and extrapolate the energy to the thermodynamic limit.
The energies at $\nu^{(1)}=1/2$ have a complicated dependence on $1/N$, which makes extrapolation to the thermodynamic limit difficult.
The following conclusions can be drawn.

(i) For all fractions shown in Fig.~\ref{fs}, the hollow-core state has a very high energy and is therefore not relevant.  

(ii) The Pfaffian wave function also has higher energy than all of the CF Fermi sea states for all filling factors studied.
In particular, it has higher energy than the fully polarized CF Fermi sea ($[N]$) in the $n=1$ LL,
in contrast to GaAs where the fully polarized CF Fermi sea has higher energy\cite{Park}.
We therefore conclude that the Pfaffian state is not stabilized in either $n=0$ or $|n|=1$ Landau level in graphene.
Interestingly, for the fully polarized state, the overlaps given in Table \ref{pfaff} indicate the Pfaffian wave function is
actually a better representation of the exact Coulomb ground state at $\nu^{(1)}=1/2$ in the $n=1$ LL of graphene than it is
of  the 5/2 state in GaAs (for the latter, the overlaps are 0.87 and 0.84 for 8 and 10 particles, respectively\cite{overlap});
nonetheless, energetic considerations rule out the Pfaffian state at $\nu^{(1)}=1/2$ in graphene.

(iii) The overlaps given in Table \ref{pfaff} show that the Pfaffian is significantly worse at $\nu^{(2)}=1/2$, indicating that it is not stabilized in the $|n|=2$ LL of graphene either.  

(iv) We have considered CF Fermi sea wave functions of four distinct symmetries, ranging from SU(4) singlet to fully polarized.
All of these have lower energies than either the Pfaffian or the hollow-core state.
Without any symmetry breaking term, the SU(4) singlet CF Fermi sea has the lowest energy at $\nu^{(0)}=1/2$,
as expected from the model of non-interacting composite fermions. When the Zeeman and the pseudo-Zeeman energies are turned on, we expect a ``partially-polarized" CF Fermi sea, and eventually a fully polarized CF Fermi sea.  

(v)  The CF Fermi sea is also favored for $\nu^{(1)}=1/2$ and $\nu^{(n)}=1/4$, but the energy differences between the various CF Fermi sea states are very small, less than the statistical error in our Monte Carlo evaluations.

\begin{table}[htb]
\begin{center}
\begin{tabular}{c|c|c|c}
\hline\hline
State & $E(\nu^{(0)}=\frac{1}{2})$ & $E(\nu^{(0)}=\frac{1}{4})$ & $E(\nu^{(1)}=\frac{1}{4})$ \\
\hline
CFFS $[N]$ & -0.4651(1)  & -0.36014(4) & n.a. \\
CFFS $[\frac{N}{2},\frac{N}{2}]$ & -0.46924(7) & -0.35955(3) & -0.3714(3) \\
CFFS $[\frac{N}{3},\frac{N}{3},\frac{N}{3}]$ & -0.4732(1)  & -0.36019(6) & -0.3720(2) \\
CFFS $[\frac{N}{4},\frac{N}{4},\frac{N}{4},\frac{N}{4}]$ & -0.47541(8) & -0.36046(6) & -0.3719(3) \\
Pfaffian & -0.45708(6) & -0.35614(2) & -0.3667(2) \\
hollow-core & -0.3141(3)  & -0.34932(3) & -0.3564(2) \\
\hline\hline
\end{tabular}
\end{center}
\caption{\label{energies} 
The thermodynamic limit of the energy per particle, in units of $e^2/\epsilon l$, for various CF Fermi sea (CFFS) states
as well as the Pfaffian and the hollow-core wave functions 
at $\nu^{(n)}=\frac{1}{2}$ and $\frac{1}{4}$ for $|n|\le1$.  The notation for the CFFS state is explained in the text.
}
\end{table}

\begin{table}[htb]
\begin{center}
\begin{tabular}{c|c|c}
\hline\hline
$N$ & $|n|=1$ & $|n|=2$ \\
\hline
 8 & 0.902 & 0.718 \\
10 & 0.894 & 0.486 \\
\hline\hline
\end{tabular}
\end{center}
\caption{\label{pfaff} 
Overlap between the Pfaffian wave function $\Psi^{\textrm{Pfaff}}_{1/2}$ and the exact
ground state (the latter obtained assuming full spin and pseudospin polarization) at $\nu^{(n)}=1/2$ in the $|n|=1,2$ Landau levels
of graphene in the spherical geometry.
(Note that the Pfaffian at $N=6$ and 12 occurs at the same flux values as $\nu=2/5$ and $3/7$, while $N=14$ is beyond our computational ability.) 
}
\end{table}

Other authors\cite{KB} have considered a CF Fermi sea state at
$\nu=0$, where the fourfold degenerate $n=0$ LL is half full.
Here, the electron (or hole) density in the $n=0$ Landau level is
$\rho=2|B|/\phi_0$, which, upon composite fermionization of {\em all}
electrons, gives an effective field of $B^\ast=|B|-2\phi_0\rho =-3|B|$
for composite fermions, which should be contrasted with $B^*=0$ at
$\nu^{(n)}=1/2$. Khveschenko\cite{KB} considers a state in which each  
of the
four degenerate Landau bands is half filled forming a CF Fermi sea; the
flux attachment does not introduce correlations between different  
bands in this approach.
Finally, we comment on some of the approximations made in the model
considered above.  We have neglected LL mixing in our calculation;
given that the energy difference between the CFFS and the Pfaffian states is fairly large (~3-5\%), we believe that LL mixing will not cause a phase transition into a Pfaffian ground state, which is known to become worse with LL mixing\cite{Wojs06}.

\section{Acknowledgements}

We thank the High Performance Computing (HPC) group at Penn State University ASET (Academic Services and Emerging Technologies)
for assistance and computing time on the Lion-XO cluster,
and the Center for Scientic Computing at J.~W.~Goethe-Universit\"at for computing time on Cluster III.

\newcommand{\PRL}{Phys.\ Rev.\ Lett.}
\newcommand{\PRB}{Phys.\ Rev.\ B}
\newcommand{\NPB}{Nucl.\ Phys.\ B}


\begin{thebibliography}{99}

\bibitem{Tsui82} D.~C.~Tsui, H.~L.~Stormer, and A.~C.~Gossard, \PRL\ {\bf 48}, 1559 (1982).

\bibitem{FQHEgraphene} C.~T\H oke, P.~E.~Lammert, V.~H.~Crespi, and J.~K.~Jain, \PRB\ \textbf{74}, 235417 (2006);
V.~M.~Apalkov and T.~Chakraborty, \PRL\ \textbf{97}, 126801 (2006).

\bibitem{graphenesu4} C.~T\H oke and J.~K.~Jain, Phys. Rev. B {\bf 75}, 245440 (2007).

\bibitem{CF} J.~K.~Jain, \PRL\ {\bf 63}, 199 (1989).

\bibitem{holes} The term ``holes" will refer in this paper to empty states in an otherwise full Landau level of graphene 
(and {\em not} to missing electrons below the $B=0$ crossing point of graphene band structure).

\bibitem{FStheory} V.~Kalmeyer and S.C.~Zhang, \PRB\ {\bf 46}, R9889 (1992); B.I.~Halperin, P.A.~Lee, and N.~Read, \PRB\ {\bf 47}, 7312 (1993).

\bibitem{FSexp} R.L.~Willett {\em et al.}, \PRL\ {\bf 71}, 3846 (1993); W.~Kang {\em et al.}, \PRL\ {\bf 71}, 3850 (1993); V.J.~Goldman {\em et al.}, \PRL\ {\bf 72}, 2065 (1994); J.H.~Smet {\em et al.}, \PRL\ {\bf 77}, 2272 (1996).

\bibitem{Park} K.~Park,V.~Melik-Alaverdian, N.~E.~Bonesteel and J.~K.~Jain, Phys. Rev. B {\bf 58}, R10167 (1998).

\bibitem{Park2} K.~Park and J.K ~Jain, \PRL\ {\bf 80}, 4237 (1998).

\bibitem{MRGWW} G.~Moore and N.~Read, \NPB\ \textbf{360}, 362 (1991);
M.~Greiter, X.~G.~Wen, and F.~Wilczek, \PRL\ \textbf{66}, 3205 (1991); \NPB\ \textbf{374}, 567 (1992).

\bibitem{HR} F.~D.~M.~Haldane and E.~H.~Rezayi, \PRL\ \textbf{60}, 956 (1988).

\bibitem{Shayegan} O.~Gunawan {\em et al.}, \PRL\ {\bf 97}, 186404 (2006).

\bibitem{Semenoff} G.~W.~Semenoff, \PRL\ \textbf{53}, 2449 (1984); F.~D.~M.~Haldane, \PRL\ \textbf{61}, 2015 (1988);
D.~P.~DiVincenzo and E.~J.~Mele, \PRB\ {\bf 29}, 1685 (1984); N.~H.~Shon and T.~Ando, J.\ Phys.\ Soc.\ Jpn.\ {\bf 67}, 2421 (1998).

\bibitem{Haldane} F.~D.~M.~Haldane, \PRL\ \textbf{51}, 605 (1983);
also in \textit{The Quantum Hall Effect}, edited by S.M.~Girvin (Springer, New York, 1987).

\bibitem{Nomura} K.~Nomura and A.~H.~MacDonald, \PRL\ \textbf{96}, 256602 (2006).

\bibitem{Fano} G.~Fano, F.~Ortolani, and E.~Colombo, \PRB\ \textbf{34}, 2670 (1986).

\bibitem{projection} J.~K.~Jain and R.~K.~Kamilla, Int.\ J.\ Mod.\ Phys.\ \textbf{B11}, 2621 (1997); \PRB\ \textbf{55}, R4895 (1997);
G.~M\"oller and S.~H.~Simon, \PRB\ \textbf{72}, 045344 (2005).

\bibitem{Willett1} R.\ Willett {\em et al.}, \PRL\ \textbf{59}, 1776 (1987);
J.~P.~Eisenstein {\em et al.}, {\em ibid.} \textbf{61}, 997 (1988); W. Pan {\em et al.}, {\em ibid.} {\bf 83}, 3530 (1999).

\bibitem{back} The physically relevant energy is obtained by extrapolation of the finite system energies  to the thermodynamic limit.
A uniformly charged positive background is assumed for a meaningful extrapolation, so that, in the spherical geometry, the total energy is given by 
\[
\left\langle\Psi_{\textrm{t}}\left|\sum_{i<j}V(\vec r_i-\vec r_j)\right|\Psi_{\textrm{t}}\right\rangle - \frac{N^2}{2R}
\]
for all trial wave functions $\Psi_{\textrm t}$, where $R$ is the radius of the sphere.  The energy differences between 
various states do not depend on the details of the background subtraction. 
Note that the graphene sheet in itself is not neutral, and part of the neutralizing charge resides on the backgate in the experimental setup;
the capacitive energy of the graphene sheet-backgate system is not taken into account, but, again, it is the same for all states, so does not change their energy ordering.

\bibitem{overlap} V.W.~Scarola, J.K.~Jain, and E.H.~Rezayi, \PRL\ {\bf 88}, 21684 (2002).

\bibitem{KB} D.~V.~Khveshchenko, \PRB\ \textbf{75}, 153405 (2007); G.~Baskaran, cond-mat/0702420 (2007).

\bibitem{Wojs06} A. W\'{o}js and J.J. Quinn, Phys. Rev. B {\bf 74}, 235319 (2006).

\bibitem{Goerbig} M. O. Goerbig, R. Moessner, and B. Dou\c{c}ot, Phys. Rev. B {\bf 74}, 161407 (2006).

\end{thebibliography}
\end{document}